\documentclass[12pt]{article}
\usepackage[dvips]{graphicx}
\usepackage{pdproc}

  \makeatletter 
  \def\@cite#1{[#1]} 
  \makeatother    
\begin{document}

\renewcommand{\thefootnote}{\alph{footnote}}

\title{
 Bounds on the Gluino Mass from a Global Parton Density Analysis 
\footnote{Argonne report ANL-HEP-CP-04-75. To be published in the 
Proceedings of the 12th International 
Conference on Supersymmetry and Unification of Fundamental Interactions
(SUSY 2004) June 17-23, 2004, Tsukuba, Japan}
}

\author{ EDMOND L. BERGER}

\address{ 
High Energy Physics Division, Argonne National Laboratory \\
Argonne, Illinois 60439, USA
%%%%% You may comment out the e-mail address line below.  
\\ {\rm E-mail: berger@anl.gov}}

\abstract{Parton distribution functions for protons are devised in 
which a light gluino is included along with standard model quark, 
antiquark, and gluon constituents. A global analysis of a large set 
of hadronic scattering data provides empirical constraints on 
the allowed range of the gluino mass as a function of the value of the 
strong coupling strength $\alpha_s(M_Z)$. We find that 
$m_{\widetilde{g}}>12$~GeV for the standard model world-average value 
$\alpha_{s}(M_{Z})=0.118$.  Gluino masses as small as 
$10$~GeV are admissible provided that $\alpha_s(M_Z) \ge 0.12$, about 
one standard deviation above the world-average value.  
Current hadron scattering data are insensitive to the presence of 
gluinos heavier than $\sim 100 - 150$~GeV.  
}

\normalsize\baselineskip=15pt

\section{Introduction}

Once admitted, ``light'' supersymmetric (SUSY) states 
(mass $m < 100$~GeV), influence hadron scattering processes in several 
ways. They change the evolution of the strong coupling strength 
$\alpha_s(\mu)$ as the scale $\mu$ is varied.  Second, they provide 
additional partonic degrees of freedom and share in the proton's 
momentum along with their standard model (SM) counterparts, 
altering the coupled evolution equations that govern the functional 
change of the parton distributions as momentum is varied. Third, they 
contribute to hard-scattering processes as incident partons and/or as 
produced particles.  For example, gluinos $\tilde{g}$ materialize as 
hadron jets and may increase the rate for jet production at large values 
of transverse energy $E_T$. Within the context of a global analysis of 
hadron scattering data, these three influences allow potentially strong 
constraints to be placed on the existence and masses of SUSY particles. 
In this paper, I summarize the results of a new study of bounds on the 
gluino mass from a global parton density function (PDF) 
analysis~\cite{Berger:2004mj}.  

In our study, we consider the effects of a light gluino only, and we do 
a series of fits to the full set of hadron scattering data used in the CTEQ6 
PDF analysis~\cite{Pumplin:2002vw} for various gluino masses 
$m_{\tilde{g}}$ and values of 
$\alpha_{s}(M_{Z})$.  All SM contributions are evaluated in perturbative 
QCD at next-to-leading order (NLO). The $\tilde{g}$ contributions to the 
parton density evolution and jet production cross section are evaluated 
at Born level (sufficient because $\tilde{g}(x)\ll q(x),\, g(x)$).  Data 
on hadron jet production from the Fermilab Tevatron, 
$p \bar p \rightarrow {\rm jet} + X$ at large $E_T$ are included in the fit.  
Quantitative bounds on $m_{\widetilde{g}}$ are obtained from a PDF error 
analysis similar to that of CTEQ6.  

\subsection{Role of $\alpha_s(M_Z)$}

Our conclusions on the allowed values of $m_{\tilde{g}}$ depend 
on what is assumed for $\alpha_s(M_Z)$, the value at which the evolution of 
$\alpha_s(\mu)$ is pinned, but we must and do fit data at all $\mu$.  
In our PDF analysis, we perform a series of fits to the hadron 
scattering data in which $\alpha_s(M_Z)$ is varied over the broad range  
$0.110\leq \alpha_s(M_{Z})\leq 0.150$.  We then impose the direct 
$Z$-pole two standard deviation ($2 \sigma$) bounds on $\alpha_s(M_Z)$ 
to establish bounds on $m_{\tilde{g}}$. 

A combined analysis of all $Z$-pole data within the context of the SM 
provides $\alpha_s(M_Z) = 0.1187 \pm 0.0027$~\cite{:2003ih}. The SM 
world-average obtained from a variety of determinations at different 
scales $\mu$ yields $\alpha_s(M_Z) = 0.1183 \pm 0.0027$~\cite{Bethke:2002rv}. 
Although we use these values to obtain our bounds on $m_{\tilde{g}}$, we are 
mindful that SUSY-QCD contributions can affect the extraction of $\alpha_s(\mu)$. 
A recent estimate~\cite{Luo:2003uw} finds $\alpha_s(M_Z) = 0.118 - 0.126 \pm 0.005$, 
where the variation in the central value arises from uncertainty in the size of 
SUSY-QCD corrections to the $Z$ width from, e.g., 
$Z \rightarrow b {\tilde b}^* {\tilde g}/{\bar b}{\tilde b}{\tilde g}$; 
${\tilde b}$ denotes a bottom squark. As remarked later, a larger value 
of $\alpha_s(M_Z)$ permits smaller 
values of $m_{\tilde{g}}$.

\subsection{Parton distribution function (PDF) analysis}

The PDF's $f(x,Q)$ are parametrized at a starting reference scale $Q_0$ and 
evolved to all $Q > Q_0$.  The gluino distribution is generated radiatively 
from gluon splitting for $Q > m_{\tilde{g}}$. The coupled evolution equations 
in the presence of a light gluino may be found in Ref.~\cite{Berger:2004mj}.
Normalized cross sections are calculated based on the parton distributions. 
Agreement with experiment is measured by $\chi^2$. The PDF shape parameters at 
$Q=Q_0$ are varied to minimize $\chi^2$. The gluino parton density remains very 
small with respect to the SM quark and gluon densities for all accessible values 
of $Q$.  The gluinos remove momentum primarily from the gluon density, for values 
of $x > 0.05$. Since hadron jet data at large values of $E_T$ are known to probe 
the gluon density at large $x$, inclusion of these data in the fit plays a strong 
role in constraining $m_{\tilde{g}}$.  

\subsection{SUSY hard-scattering contributions}

Once SUSY particles are admitted as degrees of freedom, we must consider 
their impact on all hard scattering processes.  For jet production 
in $\bar{p} + p \rightarrow {\rm jet} +X$, we include 
${\mathcal{O}}(\alpha _{s}^{2})$ subprocesses with gluinos produced in 
the final state and either one or no gluinos in the initial state:
$q+\bar{q}\rightarrow \widetilde{g}+\widetilde{g}$, 
$g+g\rightarrow \widetilde{g}+\widetilde{g}$, 
$g+\widetilde{g}\rightarrow g+\widetilde{g}$, and 
$q+\widetilde{g}\rightarrow q+\widetilde{g}$. Contributions to 
the cross section from these processes are added to the 
${\mathcal{O}}(\alpha _{s}^{2})$ SM contributions, and 
they offset partially the loss of jet rate from the smaller gluon PDF.
In deep-inelastic scattering and in massive-lepton-pair production, the 
SM processes contribute at ${\mathcal{O}}(\alpha _{s}^{0})$, but the SUSY 
processes enter at ${\mathcal{O}}(\alpha _{s}^{1})$.  We can neglect 
${\mathcal{O}}(\alpha _{s}^{1})$ SUSY subprocesses such as 
$\gamma^* + g \rightarrow \widetilde{q}+\widetilde{g}$, where $\widetilde{q}$ 
denotes a squark,    
and ${\mathcal{O}}(\alpha _{s}^{2})$ subprocesses 
$\gamma^* + q \rightarrow q + \widetilde{g}+\widetilde{g}$.  

\section{Results}

\begin{figure}[htb]
\begin{center}
\includegraphics*[width=15cm]{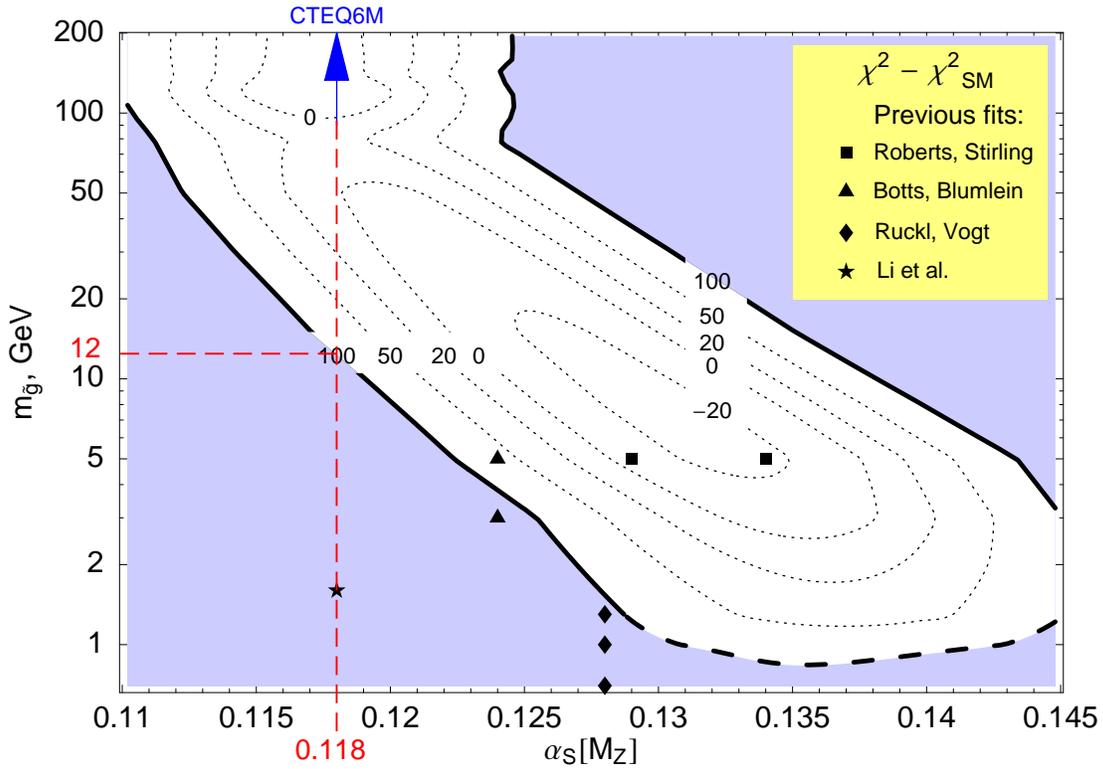}
\caption{%
Contour plot of 
$\Delta\chi^{2}=\chi^{2}\left[\alpha_{s}(M_{Z}),m_{\tilde{g}}\right]-\chi_{\mbox{CTEQ6M}}^{2}$
obtained from our fits to hadronic scattering data for various $m_{\tilde{g}}$ and 
$\alpha_{s}(M_{Z})$.  The value of $\alpha_{s}(M_{Z})$ serves as the abscissa, with the 
gluino mass along the ordinate.  The vertical dashed line marks the 
SM value $\alpha_{s}(M_{Z}) = 0.118$. }
\label{fig1}
\end{center}
\end{figure}

Our principal results are presented in the form of a contour plot.  
We show the difference $\Delta\chi^{2}$ between the value of $\chi^{2}$
obtained in our fit and that of a purely SM fit equivalent to the CTEQ6M 
fit. The CTEQ6 tolerance criterion for an acceptable fit is $\Delta\chi^{2} < 100$; 
the isoline corresponding to  
$\Delta \chi ^{2}=100$ is shown in Fig.~\ref{fig1} by the solid line. 
The acceptable fits lie inside a valley that extends from large gluino masses and
$\alpha _{s}(M_Z)=0.118$ down to $m_{\tilde{g}}\approx 0.8$~GeV and right
to $\alpha _{s}(M_Z)=0.145$.  An even narrower area corresponds to fits with 
$\chi ^{2}$ close to those in the CTEQ6M fit.  We note that $\chi ^{2}$ is 
better than in the CTEQ6M fit in a small area in which 
$m_{\tilde{g}} < 20$~GeV and $\alpha _{s}(M_{Z}) > 0.125$,
with the minimum $\Delta \chi ^{2}\approx -25$ at $m_{\tilde{g}}=8$~GeV
and $\alpha _{s}(M_Z)=0.130$. While perhaps intriguing, this negative excursion 
in $\Delta \chi ^{2}$ is smaller than the tolerance $\Delta \chi ^{2}=100$ and 
cannot be interpreted as evidence for a light gluino. We determine that 
$m_{\widetilde{g}}>12$~GeV at $\alpha_{s}(M_{Z})=0.118$.  The lower limit increases 
as $\alpha_{s}(M_{Z})$ rises; for example, with $\alpha_{s}(M_{Z})=0.122$, 
$\sim 1~\sigma$ above the SM world average, $m_{\widetilde{g}} > 5$~GeV.  

\section{Comparison with other bounds on $m_{\widetilde{g}}$}

Our global analysis is based on theoretically clean one-scale observables 
only.  We make no assumptions about the stability of the superpartners. Our 
bound $m_{\widetilde{g}}>12$~GeV is stronger than the bound 
$m_{\tilde{g}} > 6.3$~GeV at $\alpha_{s}(M_{Z})=0.118$ 
found in a study of the $Z$ boson width measurements~\cite{Janot:2003cr}. 
Negative searches for {\it stable} hadronizing gluinos provide bounds of 
$m_{\tilde{g}}>18$~GeV~\cite{Abdallah:2002qi} and 
$m_{\tilde{g}} > 26.9$~GeV~\cite{Heister:2003hc}, but these bounds do not 
apply if the gluino decays through an $R$-parity violating 
mechanism~\cite{Berger:2003kc}.  Based on a study 
of jet shape variables, the DELPHI collaboration derives a 
limit $m_{\tilde{g}} > 30-40$~GeV~\cite{Abdallah:2003xz}.
However, their analysis deals with multi-scale observables, and the treatment 
of theoretical uncertainties may be too optimistic.  
	
\section{Conclusions}

In our work, we obtain model-independent bounds on the existence and mass 
of color-octet fermions (gluinos) based on a fit to the 
complete CTEQ6 set of inclusive hadron scattering data (1811 points).  The 
data are characterized by high precision and cover a broad range in $x$ and 
$Q^2$.  We include SUSY contributions to the Tevatron jet production cross 
sections and use the full CTEQ error analysis to obtain our bounds.  
We determine that $m_{\tilde{g}}>12$~GeV for $\alpha_{s}(M_{Z})=0.118$, with  
smaller values of $m_{\tilde{g}}$ allowed for larger 
$\alpha_s(M_Z)$.  The possibility~\cite{Berger:2000mp} remains open for 
$m_{\tilde{g}} = 10 - 20$~GeV if $\alpha_{s}(M_{Z})>0.119$. For 
$\alpha_{s}(M_{Z}) >0.124$, 
$m_{\tilde{g}}$ is also bounded from above. The PDF analysis of current 
hadron scattering data is not sensitive to gluinos with mass 
$m_{\tilde{g}}$ above the weak-scale, but inclusion of a gluino with 
$m_{\tilde{g}} \sim 100$~GeV slightly improves the description of jet 
data at high $E_T$~\cite{Berger:2004mj}.    
Our study is complementary to those in which bounds $m_{\tilde{g}}$ are 
derived from LEP data, and it demonstrates the potential for PDF analysis 
to independently constrain new physics in the next few years once high 
precision data extend to larger values of $Q$.  

\section{Acknowledgments}

It is a pleasure to acknowledge the fruitful collaborative work with 
Pavel Nadolsky, Fredrick Olness, and Jon Pumplin that forms the basis 
for this paper.  Research in the High Energy Physics Division at Argonne 
National Laboratory is supported by the U.~S.~Department of Energy, 
Division of High Energy Physics, under Contract W-31-109-ENG-38.

\bibliographystyle{plain}

\begin{thebibliography}{99}
%
%\cite{Berger:2004mj}
\bibitem{Berger:2004mj}
E.~L.~Berger, P.~M.~Nadolsky, F.~I.~Olness, and J.~Pumplin,
%``Light gluino constituents of hadrons and a global analysis of hadron
%scattering data,''
arXiv:hep-ph/0406143.
%%CITATION = HEP-PH 0406143;%%

%\cite{Pumplin:2002vw}
\bibitem{Pumplin:2002vw}
J.~Pumplin, D.~R.~Stump, J.~Huston, H.~L.~Lai, P.~Nadolsky, and W.~K.~Tung,
%``New generation of parton distributions with uncertainties from global  QCD
%analysis,''
{\it JHEP}~{\bf 0207}, 012 (2002).
%%CITATION = HEP-PH 0201195;%%

%\cite{:2003ih}
\bibitem{:2003ih}
The ALEPH, DELPHI, L3, OPAL, and SLD Collaborations
and the LEP Electroweak Working Group,
%``A combination of preliminary electroweak measurements and constraints on the
%standard model,''
arXiv:hep-ex/0312023.
%%CITATION = HEP-EX 0312023;%%

%\cite{Bethke:2002rv}
\bibitem{Bethke:2002rv}
S.~Bethke,
%``alpha(s) 2002,''
{\it Nucl.~Phys.~Proc.~Suppl.}~{\bf 121}, 74 (2003).
%%CITATION = HEP-EX 0211012;%%

%\cite{Luo:2003uw}
\bibitem{Luo:2003uw}
Z.~Luo and J.~L.~Rosner,
%``Heavy bottom squark mass in the light gluino and light bottom squark
%scenario,''
{\it Phys.~Lett.~B}~{\bf 569}, 194 (2003).
%%CITATION = HEP-PH 0306022;%%

%\cite{Janot:2003cr}
\bibitem{Janot:2003cr}
P.~Janot,
%``The light gluino mass window revisited,''
{\it Phys.~Lett.~B}~{\bf 564}, 183 (2003).
%%CITATION = HEP-PH 0302076;%%

%\cite{Abdallah:2002qi}
\bibitem{Abdallah:2002qi}
J.~Abdallah {\it et al.}  [DELPHI Collaboration],
%``Search for an LSP gluino at LEP with the DELPHI detector,''
{\it Eur.~Phys.~J.~C}~{\bf 26}, 505 (2003).
%%CITATION = HEP-EX 0303024;%%

%\cite{Heister:2003hc}
\bibitem{Heister:2003hc}
A.~Heister {\it et al.}  [ALEPH Collaboration],
%``Search for stable hadronizing squarks and gluinos in e+ e- collisions  up to
%s**(1/2) = 209-GeV,''
{\it Eur.~Phys.~J.~C}~{\bf 31}, 327 (2003).
%%CITATION = HEP-EX 0305071;%%

%\cite{Berger:2003kc}
\bibitem{Berger:2003kc}
E.~L.~Berger and Z.~Sullivan,
%``Lower limits on R-parity violating couplings in supersymmetry,''
{\it Phys. Rev. Lett.}~{\bf 92}, 201801 (2004).  
%%CITATION = HEP-PH 0310001;%%

%\cite{Abdallah:2003xz}
\bibitem{Abdallah:2003xz}
J.~Abdallah {\it et al.}  [DELPHI Collaboration],
%``A study of the energy evolution of event shape distributions and their means
%with the DELPHI detector at LEP,''
{\it Eur.~Phys.~J.~C}~{\bf 29}, 285 (2003).
%%CITATION = HEP-EX 0307048;%%

%\cite{Berger:2000mp}
\bibitem{Berger:2000mp}
E.~L.~Berger, B.~W.~Harris, D.~E.~Kaplan, Z.~Sullivan, T.~M.~P.~Tait, and C.~E.~M.~Wagner,
%``Low energy supersymmetry and the Tevatron bottom-quark cross section,''
{\it Phys.~Rev.~Lett.}~ {\bf 86}, 4231 (2001).
%%CITATION = HEP-PH 0012001;%%
%
\end{thebibliography}

\end{document}